\documentclass[a4paper,10pt]{article}
\usepackage[english]{babel}
\usepackage[T1]{fontenc}
\usepackage{mathptmx}
\usepackage[cp1251]{inputenc}
\usepackage{ifpdf}
\ifpdf
 \usepackage[pdftex]{hyperref}
\else
 \usepackage[hypertex]{hyperref}
\fi
\usepackage{amsmath,amssymb,amsfonts,bm,euscript,amscd,ae,amsthm,exscale,amsopn,amstext}
\usepackage{ifthen}
\usepackage[active]{srcltx}

\title{Reduced Quantum General Relativity in Higher Dimensions}

\author{Lukasz Andrzej Glinka\footnote{\emph{Independent Science Editor \& Author, Non-fiction \& Philosophy Writer, Poland}; \emph{Member, American Association of International Researchers, USA}; E-mail: laglinka@gmail.com}~~and Patrick Linker\footnote{
\emph{M.Sc. Student, Fachgebiet Kontinuumsmechanik, Technische Universit\"at Darmstadt, 64287 Darmstadt, Germany};E-mail: patrick.linker@t-online.de}}
			
\begin{document}

\date{\today}

\maketitle

\begin{abstract}
The higher dimensional Quantum General Relativity of a Riemannian manifold being an embedded space in a space-time being a Lorentzian manifold is investigated. The model of quantum geometrodynamics, based on the Wheeler-DeWitt equation reduced to a first order functional quantum evolution supplemented through an additional eigenequation for the scalar curvature, is formulated. Furthermore, making use of the objective quantum gravity and global one-dimensional conjecture, the general wave function beyond the Feynman path integral technique is derived. The resulting quantum gravity model creates the opportunity of potentially new theoretical and phenomenological applications for astrophysics, cosmology, and physics.
\end{abstract}

\section{Introduction}\label{sec:intro}

Quantum geometrodynamics is the pioneering formulation of the quantum theory of gravitational field, for which the classical theory is the General Theory of Relativity, Cf. e.g. the Refs. \cite{01}-\cite{13}, having straightforward applications in astrophysics, cosmology, and physics, Cf. e.g. the Refs. \cite{14}-\cite{23}. Its basic standpoint is the canonical primary quantization in the Dirac Hamiltonian formalism, Cf. e.g. the Refs. \cite{24}-\cite{25}, applied to the Einstein-Hilbert action with possible inclusion of the York-Gibbons-Hawking boundary action in accordance with the Arnowitt-Deser-Misner decomposition of a space-time metric being a solution to the Einstein field equations. This strategy leads to the model of quantum gravity given through the Wheeler-DeWitt equation, the specific variant of the Schr\"odinger wave equation of quantum mechanics, for whose the dynamical object is geometry of a space embedded in a space-time and for whose the configuration space is the Wheeler superspace equipped with the DeWitt supermetric. In the present-day state of affairs, quantum geometrodynamics, as the core model for either critique or development, lays the foundations for a number models of quantum cosmology and quantum gravity and their applications, the best possible example are the professional books which were either published or republished in the recent fifteen years, Cf. e.g. the Refs. \cite{31}-\cite{40}. In this brief article, we shall consider the quantum geometrodynamics in the way which results in reduction of the Wheeler-DeWitt equation to the system of two equations: a first order quantum evolution and a supplementary eigenequation. Furthermore, we apply the objective quantum gravity and the global one-dimensional conjecture in order to construct the general solutions beyond the Feynman path integral technique.

\section{Wheeler-DeWitt Equation}

Let us consider a $1+D$-dimensional space-time being a Lorentzian manifold characterized through $(1+D)\times(1+D)$ metric tensor $g_{\mu\nu}$ of signature $(1,D)$, the Ricci curvature tensor $~{}^{(1+D)}\!R_{\mu\nu}$ and the Ricci scalar curvature $~{}^{(1+D)}\!R=g^{\mu\nu}~{}^{(1+D)}\!R_{\mu\nu}$, where a Greek index runs from $0$ to $D$. Let a $D$-dimensional embedded space being a Riemannian manifold is characterized through $D\times D$ metric tensor $h_{ij}$, the Ricci curvature tensor $~{}^{(D)}\!R_{ij}$ and the Ricci scalar curvature $~{}^{(D)}\!R=h^{ij}~{}^{(1+D)}\!R_{ij}$, where a Latin index runs from $1$ to $D$. The foundational differential geometry formulas for 4-dimensional space-time curvatures remain unchanged in higher dimensions until application of the identity $g^{\mu\nu}g_{\mu\nu}=D+1$. The Einstein tensor $~{}^{(1+D)}\!G_{\mu\nu}=~{}^{(1+D)}\!R_{\mu\nu}-\dfrac{1}{2}g_{\mu\nu}~{}^{(1+D)}\!R$, in presence of a cosmological constant $\Lambda_D$ and matter fields described through a stress-energy tensor $T_{\mu\nu}$, gives the field equations $~{}^{(1+D)}\!G_{\mu\nu}+\Lambda_D g_{\mu\nu}=\kappa_{D}T_{\mu\nu}$ where $\kappa_D=\kappa\ell_P^{D-3}$ and $\kappa=\dfrac{8\pi G}{c^4}$ is the Einstein constant. By virtue of the intrinsic geometry, the field equations for a Lorentzian manifold transform into classical geometrodynamics of a Riemannian manifold obeying the Gauss-Codazzi-Ricci equations.

\subsection{Vacuum Gravitational Fields}

The main object of quantum geometrodynamics is a wave function $\left|\Psi[h_{ij}]\right\rangle$ which is considered as a physical state. For vacuum gravitational fields, the Wheeler-DeWitt equation has the following form
\begin{equation}\label{2}
\left(G_{ijkl}\dfrac{\delta^2}{\delta h_{ij}\delta h_{kl}}+\ell_P^2\sqrt{h}~{}^{(D)}\!R\right)\left|\Psi[h_{ij}]\right\rangle=0,
\end{equation}
where $h=\det h_{ij}$, $G_{ijkl}=\dfrac{1}{2\sqrt{h}}\left(h_{ik}h_{jl}+h_{il}h_{jk}-h_{ij}h_{kl}\right)$ is the DeWitt supermetric on the Wheeler superspace, and $\ell_P=\sqrt{\dfrac{\hbar G}{c^3}}$ is the Planck length. We shall consider a special situation
\begin{equation}\label{3}
\ell_P^2{}^{(D)}\!R\left|\Psi[h_{ij}]\right\rangle=\delta^2\left|\Psi[h_{ij}]\right\rangle,
\end{equation}
with the following definition of the first variation of a wave function
\begin{equation}\label{5}
\delta \left|\Psi[h_{ij}]\right\rangle=\epsilon_{ij}\dfrac{\delta}{\delta h_{ij}}\left|\Psi[h_{ij}]\right\rangle,
\end{equation}
where the constant multiplier has been chosen for convenience and $\epsilon_{ij}$ are tensor coefficients, or equivalently
\begin{equation}\label{5a}
\delta \left|\Psi[h_{ij}]\right\rangle=2i\kappa_D\ell_P\epsilon_{ij}\pi^{ij}\left|\Psi[h_{ij}]\right\rangle,
\end{equation}
where $\pi^{ij}=-i\dfrac{1}{2\kappa_D\ell_P}\dfrac{\delta}{\delta h_{ij}}=-i\dfrac{\hslash c}{16\pi\ell_P^{D}}\dfrac{\delta}{\delta h_{ij}}$ is the canonically conjugate momentum operator in the Wheeler-Schr\"odinger representation. Calculating the second variation, one obtains
\begin{equation}\label{6}
\delta^2\left|\Psi[h_{ij}]\right\rangle=\ell_P^2{}^{(D)}\!R\left|\Psi[h_{ij}]\right\rangle=\delta\epsilon_{ij}\dfrac{\delta}{\delta h_{ij}}\left|\Psi[h_{ij}]\right\rangle+\epsilon_{ij}\epsilon_{kl}\dfrac{\delta^2}{\delta h_{ij}\delta h_{kl}}\left|\Psi[h_{ij}]\right\rangle,
\end{equation}
or, equivalently,
\begin{equation}\label{6a}
\delta^2\left|\Psi[h_{ij}]\right\rangle=\ell_P^2{}^{(D)}\!R\left|\Psi[h_{ij}]\right\rangle=2i\kappa_D\ell_P\epsilon_{ij}\pi^{ij}\left|\Psi[h_{ij}]\right\rangle-4\kappa_D^2\ell_P^2\epsilon_{ij}\epsilon_{kl}\pi^{ij}\pi^{kl}\left|\Psi[h_{ij}]\right\rangle.
\end{equation}
Whenever the relation (\ref{6}) is rewritten in the form of a wave equation
\begin{equation}
\left(-\sqrt{h}\epsilon_{ij}\epsilon_{kl}\dfrac{\delta^2}{\delta h_{ij}\delta h_{kl}}-\sqrt{h}\delta\epsilon_{ij}\dfrac{\delta}{\delta h_{ij}}+\ell_P^2\sqrt{h}~{}^{(D)}\!R\right)\left|\Psi[h_{ij}]\right\rangle=0,
\end{equation}
then it coincides with the Wheeler-DeWitt equation if
\begin{equation}\label{7}
 \delta\epsilon_{ij}\dfrac{\delta}{\delta h_{ij}}\left|\Psi[h_{ij}]\right\rangle=0,
\end{equation}
what expresses the orthogonality condition $\delta \epsilon_{ij}\pi^{ij}\left|\Psi[h_{ij}]\right\rangle=0$ and leads to the conclusion ${}^{(D)}\!R\left|\Psi[h_{ij}]\right\rangle=0$. Moreover, one this compatibility also requires $-\sqrt{h}\epsilon_{ij}\epsilon_{kl}=G_{ijkl}$ what gives the equation
\begin{equation}\label{x}
\epsilon_{ij}\epsilon_{kl}=\dfrac{1}{2h}\left(h_{ij}h_{kl}-h_{ik}h_{jl}-h_{il}h_{jk}\right),
\end{equation}
which, through contractions with space metric, gives the results
\begin{eqnarray}
\epsilon_{ij}&=&\sqrt{\dfrac{D-2}{2Dh}}h_{ij},\\
G_{ijkl}&=&-\dfrac{D-2}{2D\sqrt{h}}h_{ij}h_{kl}=-\dfrac{D-2}{2\sqrt{h}}h_{i(k}h_{jl)},
\end{eqnarray}
where the brackets refer to symmetrization. Calculating the first variation
\begin{equation}
 \delta\epsilon_{ij}=\epsilon_{kl}\dfrac{\delta}{\delta h_{kl}}\epsilon_{ij},
\end{equation}
and making slightly tedious calculations with help of the Jacobi formula $\delta h=hh^{ij}\delta h_{ij}$, one obtains
\begin{eqnarray}
\delta\epsilon_{ij}&=&-\dfrac{(D-2)^2}{4Dh}h_{ij},\\
\delta G_{ijkl}&=&2\dfrac{D(D-4)\sqrt{2D(D-2)}}{(D-2)^3}h\delta\epsilon_{ij}\delta\epsilon_{kl},
\end{eqnarray}
and, therefore, the aforementioned orthogonality condition can be expressed as $h_{ij}\pi^{ij}\left|\Psi[h_{ij}]\right\rangle=0$.

\subsection{Non-Vacuum Gravitational Fields}

Non-vacuum gravitational fields are described the Wheeler-DeWitt equation in the form
\begin{equation}\label{2a}
\left(G_{ijkl}\dfrac{\delta^2}{\delta h_{ij}\delta h_{kl}}+\ell_P^2\sqrt{h}\left(~{}^{(D)}\!R-2\Lambda_D-2\kappa_D\rho\right)\right)\left|\Psi[h_{ij},\phi]\right\rangle=0,
\end{equation}
where the symbol $\phi$ denotes dependence on matter fields, and $\rho=T_{\mu\nu}n^\mu n^\nu$ represents matter fields energy density related to a vector field $n^\mu$ normal to an embedded space. Inclusion of the strategy of the previous subsection gives
\begin{equation}\label{2b}
\left(G_{ijkl}\dfrac{\delta^2}{\delta h_{ij}\delta h_{kl}}+\sqrt{h}\left(\delta\epsilon_{ij}\dfrac{\delta}{\delta h_{ij}}+\epsilon_{ij}\epsilon_{kl}\dfrac{\delta^2}{\delta h_{ij}\delta h_{kl}}-2\ell_P^2(\Lambda_D+\kappa_D\rho)\right)\right)\left|\Psi[h_{ij},\phi]\right\rangle=0,
\end{equation}
and, through compatibility with the case of vacuum gravitational fields, consistency holds for
\begin{equation}\label{pop}
h_{ij}\dfrac{\delta}{\delta h_{ij}}\left|\Psi[h_{ij},\phi]\right\rangle=-\dfrac{8D\ell_P^2h}{(D-2)^2}\left(\Lambda_D+\kappa_D\rho\right)\left|\Psi[h_{ij},\phi]\right\rangle,
\end{equation}
or, equivalently,
\begin{equation}
ih_{ij}\pi^{ij}\left|\Psi[h_{ij},\phi]\right\rangle=-\dfrac{4D\ell_Ph}{\kappa_D(D-2)^2}\left(\Lambda_D+\kappa_D\rho\right)\left|\Psi[h_{ij},\phi]\right\rangle,
\end{equation}
and this is the reduced wave equation of quantum geometrodynamics. In this state of affairs, one can derive immediately
\begin{eqnarray}
&&h_{ij}h_{kl}\dfrac{\delta^2}{\delta h_{ij}\delta h_{kl}}\left|\Psi[h_{ij},\phi]\right\rangle=\\
&=&-\dfrac{8D^2\ell_P^2h^2}{(D-2)^2}\left[
\dfrac{\kappa_D}{Dh}h_{ij}\dfrac{\delta\rho}{\delta h_{ij}}+\Lambda_D+\kappa_D\rho
-\dfrac{8\ell_P^2}{(D-2)^2}\left(\Lambda_D+\kappa_D\rho\right)^2\right]\left|\Psi[h_{ij},\phi]\right\rangle,\nonumber
\end{eqnarray}
where we have employed the Jacobi formula, and, for this reason, one can construct the equation
\begin{eqnarray}
\!\!\!\!\!\!\!\!\!\!\!\!\!\!\!\!\!\!\!\!&&{}^{(D)}\!R\left|\Psi[h_{ij}]\right\rangle=\\
\!\!\!\!\!\!\!\!\!\!\!\!\!\!\!\!\!\!\!\!&&\dfrac{4Dh}{D-2}\left[\left(\dfrac{D-2}{2Dh}-1\right)\left(\Lambda_D+\kappa_D\rho\right)
+\dfrac{8\ell_P^2}{(D-2)^2}\left(\Lambda_D+\kappa_D\rho\right)^2-
\dfrac{\kappa_D}{Dh}h_{ij}\dfrac{\delta\rho}{\delta h_{ij}}\right]\left|\Psi[h_{ij},\phi]\right\rangle,\nonumber
\end{eqnarray}
which together with the Eq. (\ref{pop}) forms the system of equations.

\section{Global One-Dimensional Objective Quantum Gravity}

Let us apply the strategy proposed in the Ref. \cite{34}. Through some general justification on the nature of a geometrodynamical wave function which particularly involves the DeWitt supposition, one can consider the solutions to the reduced quantum geometrodynamics which are objective wave functions, i.e. are dependent on at most the matrix invariants $c_n$, $1\leqslant n\leqslant D$, of a space metric tensor $h_{ij}$. Then, a geometrodynamical wave function becomes $\left|\Psi[h_{ij}]\right\rangle=\left|\Psi(c_1,\ldots,c_D)\right\rangle$, and, moreover, one can write $\dfrac{\delta}{\delta h_{ij}}=\displaystyle\sum_{n=1}^D\dfrac{\delta c_n}{\delta h_{ij}}\dfrac{\delta}{\delta c_n}$. The next step is taking into account the global one-dimensional conjecture which stands that $\left|\Psi(c_1,\ldots,c_D)\right\rangle=\left|\Psi[h]\right\rangle$, and gives $\dfrac{\delta}{\delta c_n}=\dfrac{\delta h}{\delta c_n}\dfrac{\delta}{\delta h}$. Involving once again the Jacobi formula and the obvious identity $\displaystyle\sum_{n=1}^D\dfrac{\delta c_n}{\delta h_{ij}}\dfrac{\delta h_{ij}}{\delta c_n}=D$, this strategy leads to the relation $\dfrac{\delta}{\delta h_{ij}}=Dhh^{ij}\dfrac{\delta}{\delta h}$, and, consequently, $h_{ij}\dfrac{\delta}{\delta h_{ij}}=D^2h\dfrac{\delta}{\delta h}$ and $h_{ij}h_{kl}\dfrac{\delta^2}{\delta h_{ij}\delta h_{kl}}=D^4h^2\dfrac{\delta^2}{\delta h^2}$. In this state of affairs, the reduced wave equation derived in the previous section transforms into the objective quantum gravity
\begin{equation}\label{pop1}
\dfrac{\delta}{\delta h}\left|\Psi[h_{ij},\phi]\right\rangle=-\dfrac{8\ell_P^2}{D(D-2)^2}\left(\Lambda_D+\kappa_D\rho\right)\left|\Psi[h_{ij},\phi]\right\rangle,
\end{equation}
which also applies to vacuum gravitational fields as $\dfrac{\delta}{\delta h}\left|\Psi[h]\right\rangle=0$. Since the received model of objective quantum gravity formally is a first-order ordinary differential equation, its solution can be straightforwardly constructed avoiding the Feynman path integral method, which for the Wheeler superspace formalism is a slightly troublesome technique, in favour of a standard Cauchy problem for an initial wave function $\left|\Psi[h_I]\right\rangle$, where $h_I$ is an initial value of $h$. The general solution is an objective global one-dimensional wave function
\begin{equation}
\left|\Psi[h]\right\rangle=N\exp\left\{-\dfrac{8\ell_P^2}{D(D-2)^2}\left(\Lambda_D(h-h_I)+\kappa_D\int_{h_I}^h\rho[h']\delta h'\right)\right\}\left|\Psi[h_I]\right\rangle,
\end{equation}
where $N$ stands for a normalization factor. Remarkably, an initial state $\left|\Psi[h_I]\right\rangle$ is a solution to the vacuum theory, and in itself is undetermined. Through the global one-dimensional nature of the model, normalization can be adopted in the standard form of the Born condition $\displaystyle\int_{h_I}^{h_F} \left\langle\Psi[h']\right.\!\!\left|\Psi[h']\right\rangle\delta h'=1$ with $h_F$ being a final value of $h$ and $\left\langle\Psi[h_I]\right.\!\!\left|\Psi[h_I]\right\rangle=1$, what leads to the conclusion
\begin{equation}
N=\left\{\int_{h_I}^{h_F}\exp\left[-\dfrac{16\ell_P^2}{D(D-2)^2}\left(\Lambda_D(h'-h_I)+\kappa_D\int_{h_I}^{h'}\rho[h'']\delta h''\right)\right]\delta h'\right\}^{-1/2}.
\end{equation}
Furthermore, one can determine
\begin{equation}
\delta^2\left|\Psi[h]\right\rangle=\ell_P^2{}^{(D)}\!R\left|\Psi[h_{ij}]\right\rangle
=-\dfrac{D(D-2)^2}{4}\dfrac{\delta}{\delta h}\left|\Psi[h]\right\rangle+\dfrac{D^3(D-2)}{2}h\dfrac{\delta^2}{\delta h^2}\left|\Psi[h]\right\rangle,
\end{equation}
and calculating explicitly
\begin{equation}
\dfrac{\delta^2}{\delta h^2}\left|\Psi[h]\right\rangle
=\left[\left(\dfrac{8\ell_P^2}{D(D-2)^2}\right)^2\left(\Lambda_D+\kappa_D\rho\right)^2-\dfrac{8\ell_P^2}{D(D-2)^2}\kappa_D\dfrac{\delta\rho[h]}{\delta h}\right]\left|\Psi[h]\right\rangle,
\end{equation}
one obtains
\begin{equation}
{}^{(D)}\!R\left|\Psi[h]\right\rangle
=\left[2\left(\Lambda_D+\kappa_D\rho\right)+\dfrac{32D}{(D-2)^3}h\ell_P^2\left(\Lambda_D+\kappa_D\rho\right)^2
-\dfrac{4D^2}{D-2}h\kappa_D\dfrac{\delta\rho[h]}{\delta h}\right]\left|\Psi[h]\right\rangle.
\end{equation}

For the aforementioned objective quantum gravity one can derive two simplest situations. First of all, we can consider classical gravitational fields in presence of a cosmological constant and a constant energy density of matter fields, for which
\begin{eqnarray}
\left|\Psi[h]\right\rangle&=&N\exp\left\{-\dfrac{8\ell_P^2}{D(D-2)^2}\left(\Lambda_D+\kappa_D\rho_0\right)(h-h_I)\right\}\left|\Psi[h_I]\right\rangle,\\
N&=&\dfrac{\dfrac{4}{D-2}\left[\dfrac{1}{D}\ell_P^2\left(\Lambda_D+\kappa_D\rho_0\right)\right]^{1/2}}{\left\{1-\exp\left[-\dfrac{16\ell_P^2}{D(D-2)^2}\left(\Lambda_D+\kappa_D\rho_0\right)(h_F-h_I)\right]\right\}^{1/2}},\\
{}^{(D)}\!R\left|\Psi[h]\right\rangle&=&
\left[2\left(\Lambda_D+\kappa_D\rho_0\right)+\dfrac{32D}{(D-2)^3}h\ell_P^2\left(\Lambda_D+\kappa_D\rho_0\right)^2\right]\left|\Psi[h]\right\rangle,
\end{eqnarray}
and two particular situations, lambda-vacuum gravitational fields and vanishing cosmological constant, can be immediately derived from this case. More general case is a flat embedded space, for which one has the objective simplex-projected state and the normalization factor in the general forms, and
\begin{equation}
\left[2\left(\Lambda_D+\kappa_D\rho\right)+\dfrac{32D}{(D-2)^3}h\ell_P^2\left(\Lambda_D+\kappa_D\rho\right)^2
-\dfrac{4D^2}{D-2}h\kappa_D\dfrac{\delta\rho[h]}{\delta h}\right]\left|\Psi[h]\right\rangle=0.
\end{equation}

\section{Discussion}

We have shown that inclusion of the reduced quantum geometrodynamics creates the model of quantum gravity based on the Wheeler-DeWitt equation. Looking through the prism of the Wheeler superspace, the aforementioned orthogonality condition related to a vacuum gravitational field has the physical interpretation as the analogue of the Lorentz/Lorenz gauge from the Maxwell electrodynamics, whereas for a general case of a non-vacuum gravitational field one obtains a considerably simplified model of quantum geometrodynamics. For this reason, the applied method has the most natural physical interpretation as the choice of a superspatial gauge. Equivalently, this technique can be regarded as the choice of the specific coordinate system in the configuration space of the General Theory of Relativity. Making use of the objective quantum gravity and the global one-dimensional conjecture for a geometrodynamical wave function, we have generated the global one-dimensional objective quantum gravity of an embedded Riemannian manifold. Remarkably, in the reduced quantum geometrodynamics all physical paths are compatible with the Wheeler-DeWitt equation, and, for this reason, the emerging model can be approached through the Feynman path integral formulation of quantum mechanics according to the idea of the Hartle-Hawking wave function. Prior to inclusion of the objective quantum gravity and the global one-dimensional conjecture, still the Feynman path integral type solutions to the Wheeler-DeWitt equation can be constructed, but inclusion of these two ideas leads to the solutions which are beyond the Feynman path integral formulation and are received through a solution to the Cauchy problem. Naturally, the most intriguing consequences of the reduced quantum general relativity, including the global one-dimensional objective quantum gravity, would be the particular phenomenological applications of the resulting model of quantum gravity. In our opinion, the scope of possible quantum gravitational effects could effect the phenomena of astrophysics, cosmology, and physics. Furthermore, pure theoretical applications of the proposed method in the context of loop quantization and string theory in themselves would be the great challenges for further development of the model.

\section*{Acknowledgements}

L.A. Glinka thanks to P. Linker for private communication which led to this joined work.

\bibliographystyle{IEEEtran}

\small
\begin{thebibliography}{99}

\bibitem{01} A. Einstein, \emph{The Meaning of Relativity: Including the Relativistic Theory of the Non-Symmetric Field}. Fifth Edition, With a New Introduction by Brian Greene. Princeton University Press, 2014.

\bibitem{02} A. Einstein, \emph{Relativity: The Special and the General Theory}. 100th Anniversary Edition, with Commentaries and Background Material by Hanoch Gutfreund and J\"urgen Renn. Princeton University Press, 2015.

\bibitem{03} M. Born, \emph{Einstein's Theory of Relativity}. Revised Edition, prepared with the collaboration of G\"unther Leibfried and Walter Biem. Dover Publications, 1965.

\bibitem{04} S. Chandrasekhar, \emph{The Mathematical Theory of Black Holes}. Oxford University Press, 2004.

\bibitem{05} P.A.M. Dirac, \emph{General Theory of Relativity}. Princeton University Press, 1996.

\bibitem{06} R.P. Feynman, \emph{Lectures on Gravitation, 1962-63}, Lecture Notes by F.B. Morinigo and W.G. Wagner. California Institute of Technology, 1963.

\bibitem{07} R.P. Feynman, \emph{Six Not-So-Easy Pieces: Einstein's Relativity, Symmetry and Space-Time}. Originally prepared for publication by Robert B. Leighton and Matthew Sands, New Introduction by Roger Penrose. Addison-Wesley, 1997.

\bibitem{08} G. 't Hooft, \emph{Introduction to General Relativity}. Rinton Press, 2001.

\bibitem{09} M. von Laue, \emph{Das Relativit\"{a}tstheorie, Zweiter Band: Die allgemeine Relativit\"atstheorie und Einsteins Lehre von der Schwerkraft}. Vieweg, 1921.

\bibitem{09a} H.A. Lorentz, \emph{The Einstein Theory of Relativity: A Concise Statement}. Brentano's, 1920.

\bibitem{10} W. Pauli, \emph{Theory of Relativity}. Revised Edition. Dover Publications, 1981.

\bibitem{11} E. Schr\"odinger, \emph{Space-Time Structure}. Cambridge University Press, 1997.

\bibitem{12} J. Schwinger, \emph{Einstein's Legacy: The Unity of Space and Time}. Dover Publications, 2002.

\bibitem{13} S. Weinberg, \emph{Gravitation and Cosmology. Principles and Applications of the General Theory of Relativity},  Wiley, 1972.

\bibitem{14} H. Alfv\'{e}n, \emph{Worlds-Antiworlds: Antimatter in Cosmology}. Freeman, 1966.

\bibitem{15a} M. Allais, \emph{L'Anisotropie de l'Espace: La N\'{e}cessaire R\'{e}vision de Certains Postulats des Th\'{e}ories Contemporaines}. Juglar, 1997.

\bibitem{15} M. Allais, \emph{De Tr\`{e}s Remarquables R\'{e}gularit\'{e}s dans les Distributions des Plan\`{e}tes et des Satellites des Plan\`{e}tes}. Juglar, 2005.

\bibitem{15b} M. Allais, \emph{L'Effondrement de la Th\'{e}orie de la Relativit\'{e}: Implication Irr\'{e}fragable des Donn\'{e}es de l'Exp\'{e}rience}. Juglar, 2004.

\bibitem{16} S. Chandrasekhar, \emph{An Introduction to the Study of Stellar Structure}. Dover Publications, 2010.

\bibitem{17} S. Chandrasekhar, \emph{Principles of Stellar Dynamics}. Dover Publications, 2005.

\bibitem{18} V.L. Ginzburg, \emph{Physics and Astrophysics: A Selection of Key Problems}. Pergamon Press, 1985.

\bibitem{19} V.L. Ginzburg, \emph{Theoretical Physics and Astrophysics}. Pergamon Press, 1969.

\bibitem{19a} V.L. Ginzburg, \emph{Waynflete Lectures on Physics: Selected Topics in Contemporary Physics and Astrophysics}. Pergamon Press, 1983.

\bibitem{20} A. Hewish, \emph{Physics of the Universe}. Publications \& Information Directorate, Council of Scientific \& Industrial Research, 1992.

\bibitem{20a} Y. Nambu, \emph{Quarks: Frontiers in Elementary Particle Physics}. World Scientific Publishing, 1985.

\bibitem{21} E. Schr\"odinger, \emph{Expanding Universes}. Cambridge University Press, 2011.

\bibitem{21a} J. Schwinger, \emph{Particles, Sources, and Fields, Volume I}. Perseus Books, 1998.

\bibitem{22} H.C. Urey, \emph{The Planets: Their Origin and Development}. Yale University Press, 1952.

\bibitem{23b} M. Veltman, \emph{Facts and Mysteries in Elementary Particle Physics}. World Scientific Publishing, 2003.

\bibitem{23a} S. Weinberg, \emph{Cosmology}. Oxford University Press, 2008.

\bibitem{23} S. Weinberg, \emph{The Quantum Theory of Fields, Volume 3: Supersymmetry}. Cambridge University Press, 2000.

\bibitem{24} P.A.M. Dirac, \emph{Lectures on Quantum Mechanics}. Belfer Graduate School of Science, Yeshiva University, 1964.

\bibitem{25} S. Weinberg, \emph{Lectures on Quantum Mechanics}. Cambridge University Press, 2013.

\bibitem{31} J. Ambj{\o}rn, B. Durhuus, and T. Jonsson, \emph{Quantum Geometry: A Statistical Field Theory Approach}. Cambridge University Press, 2005.

\bibitem{31v} M. Ammon and J. Erdmenger, \emph{Gauge/Gravity Duality: Foundations and Applications}. Cambridge University Press, 2015.

\bibitem{31u} G. Auletta, \emph{Foundations and Interpretation of Quantum Mechanics: In the Light of a Critical-historical Analysis of the Problems and of a Synthesis of the Results}. World Scientific Publishing, 2001.

\bibitem{31j} K. Becker, M. Becker, J.H. Schwarz, \emph{String Theory and M-Theory: A Modern Introduction}. Cambridge University Press, 2007.

\bibitem{31e} M. Bojowald, \emph{Canonical Gravity and Applications: Cosmology, Black Holes, Quantum Gravity}. Cambridge University Press, 2011.

\bibitem{31s} M. Bojowald, \emph{Quantum Cosmology: A Fundamental Description of the Universe}. Springer, 2011.

\bibitem{31d} M. Bojowald, \emph{The Universe: A View from Classical and Quantum Gravity}. Wiley-VCH, 2013.

\bibitem{31k} L. Brink and M. Henneaux, \emph{Principles of String Theory}. Springer, 2013.

\bibitem{31a} S. Carlip, \emph{Quantum Gravity in 2+1 Dimensions}. Cambridge University Press, 2003.

\bibitem{31b} R.W. Carroll, \emph{Fluctuations, Information, Gravity and the Quantum Potential}. Springer, 2006.

\bibitem{31c} R.W. Carroll, \emph{On the Emergence Theme of Physics}. World Scientific Publishing, 2010.

\bibitem{31r} F. Cianfrani, O.M. Lecian, M. Lulli, and G. Montani, \emph{Canonical Quantum Gravity: Fundamentals and Recent Developments}. World Scientific Publishing, 2014.

\bibitem{32} L.B. Crowell, \emph{Quantum Fluctuations of Spacetime}. World Scientific Publishing, 2005.

\bibitem{32c} P.D. D'Eath, \emph{Supersymmetric Quantum Cosmology}. Cambridge University Press, 2005.

\bibitem{32b} G. Esposito, \emph{Quantum Gravity in Four Dimensions}. Nova Science Publishers, 2001.

\bibitem{32y} G. Esposito, \emph{Quantum Gravity, Quantum Cosmology and Lorentzian Geometries}. Springer, 2013.

\bibitem{32i} G. Esposito, A.Yu. Kamenshchik, and G. Pollifrone, \emph{Euclidean Quantum Gravity on Manifolds with Boundary}. Springer, 2012.

\bibitem{32a} R. Gambini and J. Pullin, \emph{A First Course in Loop Quantum Gravity}. Oxford University Press, 2011.

\bibitem{32d} R. Gambini and J. Pullin, \emph{Loops, Knots, Gauge Theories and Quantum Gravity}. Cambridge University Press, 2000.

\bibitem{33} M. Gasperini, \emph{Elements of String Cosmology}. Cambridge University Press, 2007.

\bibitem{34} L.A. Glinka, \emph{Aethereal Multiverse: A New Unifying Theoretical Approach to Cosmology, Particle Physics, and Quantum Gravity}. Cambridge International Science Publishing, 2012.

\bibitem{35} H.W. Hamber, \emph{Quantum Gravitation: The Feynman Path Integral Approach}. Springer-Verlag, 2009.

\bibitem{35a} S. Hawking and R. Penrose, \emph{The Nature of Space and Time}. Princeton University Press, 2015.

\bibitem{35u} J.N. Isham, \emph{An Introduction to Mathematical Cosmology}. Cambridge University Press, 2002.

\bibitem{36a} C.V. Johnson, \emph{D-Branes}. Cambridge University Press, 2003.

\bibitem{36s} E. Joos, \emph{Decoherence and the Appearance of a Classical World in Quantum Theory}. Springer, 2003.

\bibitem{36} C. Kiefer, \emph{Quantum Gravity}. Third Edition. Oxford University Press, 2012.

\bibitem{36i} I. Licata and D. Fiscaletti, \emph{Quantum Potential: Physics, Geometry and Algebra}. Springer, 2013.

\bibitem{36y} D.-E. Liebscher, \emph{Cosmology}. Springer, 2005.

\bibitem{36t} A.D. Linde, \emph{Inflation and Quantum Cosmology}. Elsevier, 2012.

\bibitem{37b} P.V. Moniz, \emph{Quantum Cosmology - The Supersymmetric Perspective - Vol. 1:
Fundamentals; Vol. 2: Advanced Topic}. Springer, 2010.

\bibitem{37j} G. Montani, \emph{Primordial Cosmology}. World Scientific Publishing, 2011.

\bibitem{37} V. Mukhanov and S. Winitzki, \emph{Introduction to Quantum Effects in Gravity}. Cambridge University Press, 2007.

\bibitem{37r} J.V. Narlikar and T. Padmanabhan, \emph{Gravity, Gauge Theories and Quantum Cosmology}. Springer, 2012.

\bibitem{37a} T. Ort\'{\i}n, \emph{Gravity and Strings}. Second Edition. Cambridge University Press, 2015.

\bibitem{37h} T. Padmanabhan, \emph{Gravitation: Foundations and Frontiers}. Cambridge University Press, 2010.

\bibitem{37g} V.N. Pervushin and A. Pavlov, \emph{Principles of Quantum Universe}. Lambert Academic Publishing, 2014.

\bibitem{37p} V.N. Popov, \emph{Functional Integrals in Quantum Field Theory and Statistical Physics}. Springer, 2001.

\bibitem{38u} E. Prugove\v{c}ki, \emph{Quantum Geometry: A Framework for Quantum General Relativity}. Springer, 2013.

\bibitem{38} C. Rovelli, \emph{Quantum Gravity}. Cambridge University Press, 2004.

\bibitem{38a} C. Rovelli and F. Vidotto, \emph{Covariant Loop Quantum Gravity: An Elementary Introduction to Quantum Gravity and Spinfoam Theory}. Cambridge University Press, 2014.

\bibitem{38b} Y. Sato, \emph{Space-Time Foliation in Quantum Gravity}. Springer, 2014.

\bibitem{39} C. Simeone, \emph{Deparametrization and Path Integral Quantization of Cosmological Models}. World Scientific Publishing, 2001.

\bibitem{39f} C.F. Steinwachs, \emph{Non-minimal Higgs Inflation and Frame Dependence in Cosmology}. Springer, 2013.

\bibitem{40a} T. Thiemann, \emph{Modern Canonical Quantum General Relativity}. Cambridge University Press, 2007.

\bibitem{40b} S. Winitzki, \emph{Eternal Inflation}. World Scientific Publishing, 2009.

\bibitem{40} H.D. Zeh, \emph{The Physical Basis of The Direction of Time}. Springer, 2007.

\end{thebibliography}

\end{document}